Research

**Open Access**

# Uncovering protein interaction in abstracts and text using a novel linear model and word proximity networks

Alaa Abi-Haidar[1,2], Jasleen Kaur[1], Ana Maguitman[3], Predrag Radivojac[1], Andreas Rechtsteiner[4], Karin Verspoor[5], Zhiping Wang[6] and Luis M Rocha[1,2]

Addresses: [1]School of Informatics, Indiana University, 107 S. Indiana Ave. Bloomington, IN 47405, USA. [2]FLAD (Fundação Luso-Americana para o Desenvolvimento) Computational Biology Collaboratorium, Instituto Gulbenkian de Ciência, Rua da Quinta Grande, 6 P-2780-156 Oeiras, Portugal. [3]Departamento de Ciencias e Ingenería de la Computación, Universidad Nacional del Sur, Avenida Alem 1253, Bahía Blanca, Buenos Aires, Argentina. [4]Center for Genomics and Bioinformatics, Indiana University, 107 S. Indiana Ave. Bloomington, IN 47405, USA. [5]Modeling, Algorithms and Informatics Group, Los Alamos National Laboratory, 1350 Central, MS C330 Los Alamos, NM 87545, USA. [6]Biostatistics, School of Medicine, Indiana University, 107 S. Indiana Ave. Bloomington, IN 47405, USA.

Correspondence: Luis M Rocha. Email: rocha@indiana.edu

## Abstract

**Background:** We participated in three of the protein-protein interaction subtasks of the Second BioCreative Challenge: classification of abstracts relevant for protein-protein interaction (interaction article subtask [IAS]), discovery of protein pairs (interaction pair subtask [IPS]), and identification of text passages characterizing protein interaction (interaction sentences subtask [ISS]) in full-text documents. We approached the abstract classification task with a novel, lightweight linear model inspired by spam detection techniques, as well as an uncertainty-based integration scheme. We also used a support vector machine and singular value decomposition on the same features for comparison purposes. Our approach to the full-text subtasks (protein pair and passage identification) includes a feature expansion method based on word proximity networks.

**Results:** Our approach to the abstract classification task (IAS) was among the top submissions for this task in terms of measures of performance used in the challenge evaluation (accuracy, F-score, and area under the receiver operating characteristic curve). We also report on a web tool that we produced using our approach: the Protein Interaction Abstract Relevance Evaluator (PIARE). Our approach to the full-text tasks resulted in one of the highest recall rates as well as mean reciprocal rank of correct passages.

**Conclusion:** Our approach to abstract classification shows that a simple linear model, using relatively few features, can generalize and uncover the conceptual nature of protein-protein interactions from the bibliome. Because the novel approach is based on a rather lightweight linear model, it can easily be ported and applied to similar problems. In full-text problems, the expansion of word features with word proximity networks is shown to be useful, although the need for some improvements is discussed.

## Background

Much of the research presently conducted in genome biology relies on the induction of correlations and interactions from data. Indeed, during the past decade, fueled by the production of large biomedical databases (particularly those containing genomic data) as well as the widespread use of high-throughput technology, we have witnessed the emergence of a more data-driven paradigm for biological research, which in turn has created new analysis challenges. Because we ultimately want to increase our knowledge of the biochemical and functional roles of genes and proteins in organisms, there is a clear need to integrate the associations and interactions among biological entities that have been reported and accumulate in the literature and databases. Such integration can provide a comprehensive perspective on presently accumulated experimental knowledge, and may even uncover new relationships and interactions induced from global information but unreported in individual experiments.

Literature mining [1,2] is expected to help with such integration and inference. Its objective is to automatically sort through huge collections of literature and databases (the 'bibliome') and suggest the most relevant pieces of information for a specific analysis task, for example the annotation of proteins [3]. Given the size of the bibliome, it is no surprise that literature mining has become an important component of bioinformatics. However, this success has raised the important issue of validation and comparison of the inferences uncovered via 'bibliome informatics'. Although it is difficult to develop a 'gold standard' for all literature mining approaches, it is important to provide a means to test and validate different algorithms and electronic sources of biological knowledge. Thus, researchers in this field have focused on testing algorithms and resources on specific tasks, for instance protein annotation [3] and protein family [4] and structure [5] prediction. The BioCreative (Critical Assessment of Information Extraction systems in Biology) Challenge evaluation is precisely an effort to enable comparison of various approaches to literature mining. Perhaps its greatest value is that it consists of a community-wide effort, leading many different groups to test their methods against a common set of specific tasks, thus resulting in important benchmarks for future research.

We participated in three subtasks of the second Biocreative challenge [6], which focused on protein-protein interactions (PPIs):

1. Protein interaction article (interaction article subtask [IAS]): classify PubMed abstracts as being relevant or irrelevant to PPI. Relevance implies that articles 'are useful to derive protein interaction annotations according to the curation standards used by the' IntAct [7] and Molecular Interactions Database (MINT) [8] databases [6].

2. Protein interaction pairs (interaction pair subtask [IPS]): identify interacting protein pairs from full-text articles. Pairs of interactors are classified as such according to the curation standards of IntAct and MINT - that is, only for colocalizations and physical interactions.

3. Protein interaction sentences (interaction sentences subtask [ISS]): identify the sentences most relevant to characterize the interaction pairs identified in the IPS task.

In most text-mining projects in biomedicine, one must first collect a set of relevant documents, typically from abstract information. Such a binary classification, between relevant and irrelevant documents for PPI, is precisely what the IAS subtask in BioCreative II aimed to evaluate. Naturally, tools developed for IAS have great potential to be applied in many other text-mining projects beyond PPI. For that reason, we opted to produce a very general and lightweight system that can easily be applied to other domains and ported to different computer infrastructure. This design criteria lead us to a novel linear model inspired by spam-detection techniques. For comparison purposes, we also used a support vector machine (SVM) and singular value decomposition (SVD) enhanced with an uncertainty-based integration scheme.

As for the IPS and ISS subtasks, our approach is centered on a feature expansion method, using word proximity networks, which we introduced in the first Biocreative challenge [9]. Below, we describe our approach in detail and discuss our very positive results. We also report on a web tool we produced using our IAS approach: the Protein Interaction Abstract Relevance Evaluator (PIARE).

## Results and discussion

### Protein interaction article subtask

We submitted three runs for the IAS task of the BioCreative II challenge, each using a different abstract classification method but all using the same word and protein mention features (as described in the Materials and methods section, below). Run 1 used a SVM, run 2 used our novel variable trigonometric threshold (VTT), and run 3 used SVD with uncertainty integration (SVD-UI). The VTT and SVD-UI submissions (runs 2 and 3) to the IAS task of the BioCreative II challenge initially suffered from a software error in the computation of the feature sets used in VTT: bigrams$^+$ and abstract co-occurrence word pairs (see Materials and methods [below]). Even with this error, run 2 was still among the best performing submissions in the challenge; only 7 out of 51 other submissions (from 4 out of 19 groups) reported higher values of all three main performance measures used [10]. In [11] we discussed in detail our results from the runs originally submitted to the challenge. Here, we report on the improved results of a corrected VTT classification after fixing the software error in the computation of features, trained using - exclusively - the resources that were available at the

**Table 1**

**Performance measures of runs submitted to IAS**

| | SVM | VTT | | SVD-UI | | | | |
|---|---|---|---|---|---|---|---|---|
| | Run 1 | Run 2 | Run 2' | Run 3 | Run 3' | Mean[a] | StDev[a] | Median[a] |
| Total predictions | | | 750 | | | | | |
| Total positive | | | 375 | | | | | |
| Total negative | | | 375 | | | | | |
| True positives (TP) | 330 | 295 | 323 | 300 | | | N/A | |
| False positives (FP) | 186 | 118 | 133 | 143 | | | | |
| True negatives (TN) | 189 | 257 | 242 | 232 | | | | |
| False negatives (FN) | 45 | 80 | 52 | 75 | | | | |
| Precision | 0.64 | 0.71 | 0.71 | 0.68 | | 0.66 | 0.08 | 0.68 |
| Recall | 0.88 | 0.79 | 0.86 | 0.8 | | 0.76 | 0.19 | 0.85 |
| Accuracy | 0.69 | 0.74 | **0.75** | 0.71 | | 0.67 | 0.06 | 0.67 |
| F-score | 0.74 | 0.75 | **0.78** | 0.73 | | 0.69 | 0.10 | 0.72 |
| FP rate | 0.5 | 0.32 | 0.36 | 0.38 | | | N/A | |
| TP rate | 0.88 | 0.79 | 0.86 | 0.8 | | | | |
| Error rate | 0.31 | 0.26 | 0.25 | 0.29 | | | | |
| AUC | **0.8** | 0.76 | **0.8** | 0.71 | 0.75 | 0.74 | 0.07 | 0.75 |

[a]Calculated from 51 runs submitted by 19 teams. AUC, area under the curve; IAS, interaction article subtask; SVD, singular value decomposition; SVM, support vector machine; SVD-UI, SVD with uncertainty integration; VTT, variable trigonometric threshold. Bold entries for accuracy, F-Score, and AUC denote best value obtained for all our submitted runs.

time of the challenge. The performance of the three runs can be seen in Table 1, where run *i* denotes the runs originally submitted to the BioCreative II challenge, and Run *i*' denotes the respective corrected versions.

As can be seen in Table 1, all of our three runs were above the mean and median values of accuracy, F-score, and area under the receiver operating characteristic curve (AUC) measures computed from the results of all 51 submissions to the challenge [10]. We can also report that our novel VTT method performed better than our two other runs: SVM and SVD-UI. Moreover, the corrected VTT run improved from the submitted version; only 2 out of 51 other submissions (from 1 out of 19 groups) report higher values of all three performance measures above [10].

Let us now look in more detail at the performance according to these three measures. Accuracy gives us the ratio of correct predictions for both positive and negative abstracts (relevant and nonrelevant for protein interaction). In this case, the VTT method yielded the best result (0.75), followed by our SVD-UI (0.71) and SVM (0.71) methods. Thus, the VTT method produced a more balanced prediction for both positive and negative abstracts, leading to the lowest error rate (0.25). We should also notice that the accuracy of VTT was well above the mean plus one standard deviation for all submissions to the challenge. In fact, only 2 submissions (from the same group) out of 51 to the challenge reported a higher value of accuracy. The $F_1$ measure (or F-score) is defined as $F = 2 \times \text{precision} \times \text{recall}/(\text{precision} + \text{recall})$, where precision is the proportion of abstracts returned that are relevant (positive), and recall is the proportion of relevant abstracts that are retrieved. In this case, again VTT yielded the best result (0.78), followed by our SVM (0.74) and SVD-UI (0.73) methods. Notice that the F-score of VTT was very close to the mean plus one standard deviation for all submissions to the challenge. In fact, only 3 submissions (from 2 groups) out of 51 to the challenge reported a higher F-score. The AUC measure can be understood as the probability that for a randomly picked positive abstract and a randomly picked negative abstract, the positive abstract is ranked above the negative one for protein interaction relevance. We obtained very good results with this measure for both VTT and SVM runs (0.8), and slightly above the mean for SVD-UI (0.75). This means that the probability of finding a false positive closer to the top of the abstract rank (or a false negative closer to the bottom of the rank) produced by the SVD-UI method is considerably larger than in the ranking produced by the VTT and SVM methods. Furthermore, note that the AUC of both the VTT and SVM runs was close to the mean plus one standard deviation for all submissions to the challenge. Only 10 submissions (from 6 groups) out of 51 to the challenge reported a higher value of AUC than these two runs.

As we discuss in the Materials and methods section (below), the SVD vector model alone produced the same classification of the test abstracts as SVD-UI, except that different rankings of abstracts were attained. Therefore, the values of accuracy and F-score are identical for the SVD vector model alone and SVD-UI. However, the AUC of the SVD method alone was much lower (0.68) than that of the SVD-UI method (0.75). We can thus say that the integration method improved the AUC of the SVD method alone. On the other hand, its performance according to accuracy, F-score, and AUC was worse than the other constituent methods employed in the uncertainty integration, such as VTT as submitted in run 2. Thus, uncertainty integration did not improve the VTT alone. The fairly lackluster performance of this uncertainty integration method is possibly due to computing Shannon's entropy for the only two classes of this problem: positives and negatives. The method was originally developed [4] to classify more than 1,000 PFAM protein families, which is much more appropriate for this uncertainty measure. A probability distribution on two elements is not an ideal situation for calculating Shannon's entropy.

A comparison of all our methods in the accuracy/F-score, accuracy/AUC, and F-score/AUC planes is depicted in Figures 1, 2, and 3, respectively. The figures also contrast our results with the central tendency of all group submissions. The most salient points are as follows:

1. Accuracy: all three runs are above the mean and median values of accuracy for all teams. Run 2' (VTT) yielded an accuracy one standard deviation above the mean accuracy.

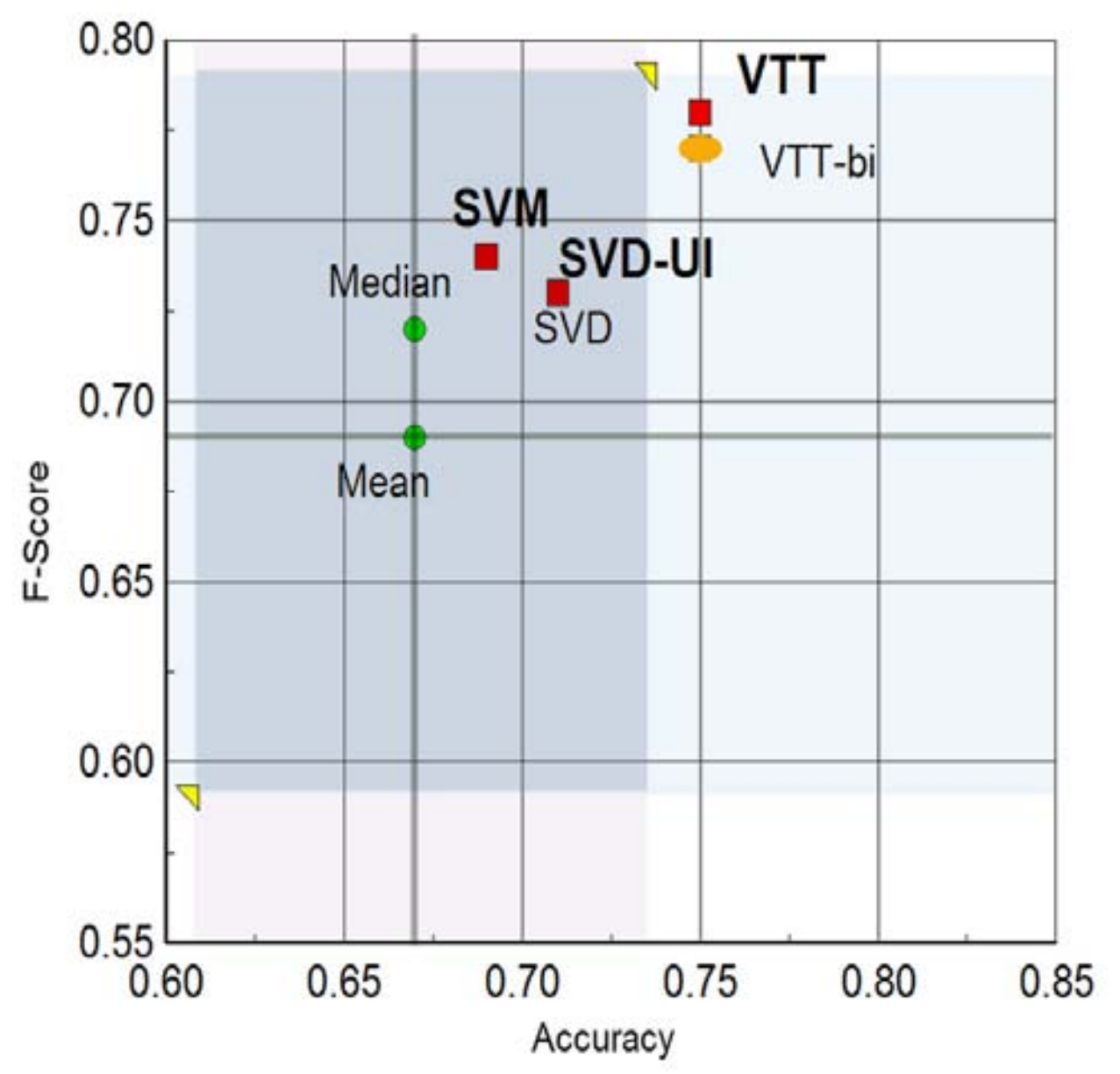

**Figure 1**
Accuracy versus F-score plane. Our methods on the accuracy versus F-score plane for IAS. Mean and median are for the set of all submissions from all groups. Red squares denote our three submissions (SVM, VTT, and SVD-UI). In this plane, SVD alone occupies the same point as SVD-UI. The orange oval denotes the results for one of the versions of VTT (with bigrams+) included in the SVD-UI method. IAS, interaction article subtask; SVD, singular value decomposition; SVM, support vector machine; SVD-UI, SVD with uncertainty integration; VTT, variable trigonometric threshold.

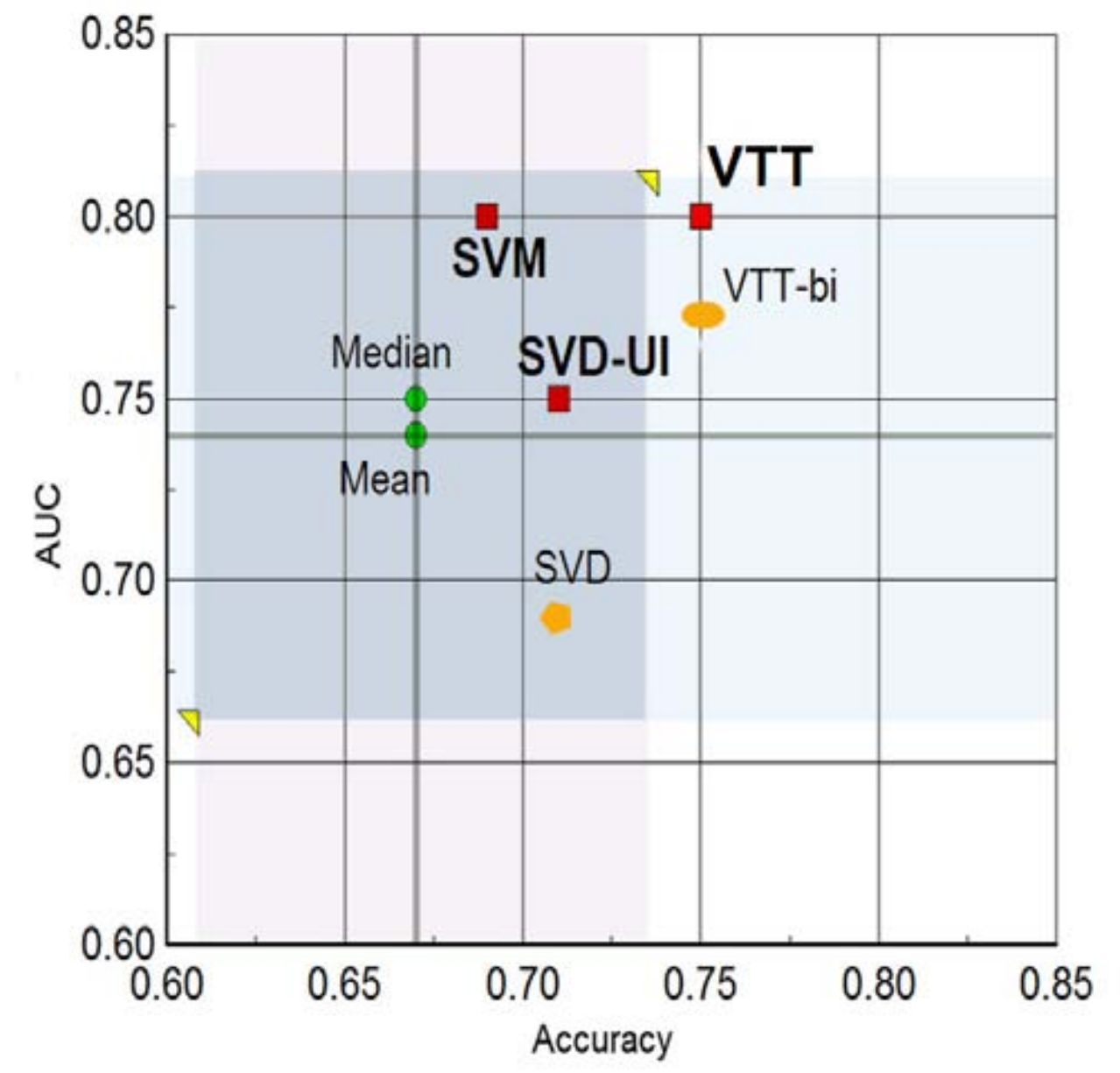

**Figure 2**
Accuracy versus AUC plane. Our methods on the accuracy versus AUC plane for IAS. Mean and median are for the set of all submissions from all groups. Red squares denote our three submissions (SVM, VTT, and SVD-UI). The orange polygon denotes the results for SVD alone, and the orange oval denotes the results for one of the versions of VTT (with bigrams+) included in the SVD-UI method. AUC, area under the receiver operating characteristic curve; IAS, interaction article subtask; SVD, singular value decomposition; SVM, support vector machine; SVD-UI, SVD with uncertainty integration; VTT, variable trigonometric threshold.

2. F-score: all three runs are above the mean and median values of F-score for all teams. Run 2' (VTT) yielded an F-score nearly one standard deviation above the mean accuracy.

3. AUC: all three runs are above the mean and median values of AUC for all teams. Runs 1 and 2' (SVM and VTT) yielded nearly one standard deviation above the mean.

4. Balance across all performance measures: the VTT method (run 2') was the only one that was quite above the mean and median for all measures tested (precision, recall, F-score, accuracy, and AUC).

### *Data issues and training*

One of the problems encountered by all methods, but particularly so for our SVM and SVD methods, was the significant difference between the training and the test IAS data in BioCreative II. It is clear that the abstracts in the training data are distinct from those in the test data. To quantify this distinction, after the challenge we trained a SVM model to classify labeled and unlabeled data - that is, between training and test data, regardless of them being relevant (positive) or irrelevant (negative) for protein interaction. If the two sets of abstracts were sampled from the same coherent semantic classes, then the AUC should be about 0.5, indicating that training and test data were indistinguishable. Instead, we obtained an AUC of 0.69. We also obtained a very high precision (0.88) and F-score (0.92). Thus, we can say that there are quite a few labeled data points (training abstracts) that are not close to anything in the unlabeled dataset (test abstracts). Because it appears possible to 'recognize' whether an abstract is from the training or the test set, one could improve classification using this information. Clearly, such an exercise would not lead to a useful system architecture for real world applications, where we do not have a 'test set'. Nonetheless, it is interesting to note that classification in the challenge can be improved because of the evident distinction between training and test data. We chose not to do this in the official submission [11], but we tested such an approach after the challenge; using the labeled versus unlabeled data classifier just described, we bootstrapped a new training set that is more similar to the test set. The bootstrapped training set is built by sampling, with replacement, a number of data points from the labeled (training) data according to the probability of belonging to the unlabeled data. This probability is obtained via the labeled versus unlabeled data classifier. The idea is that the bootstrapped training data are as close to the unlabeled data as possible.

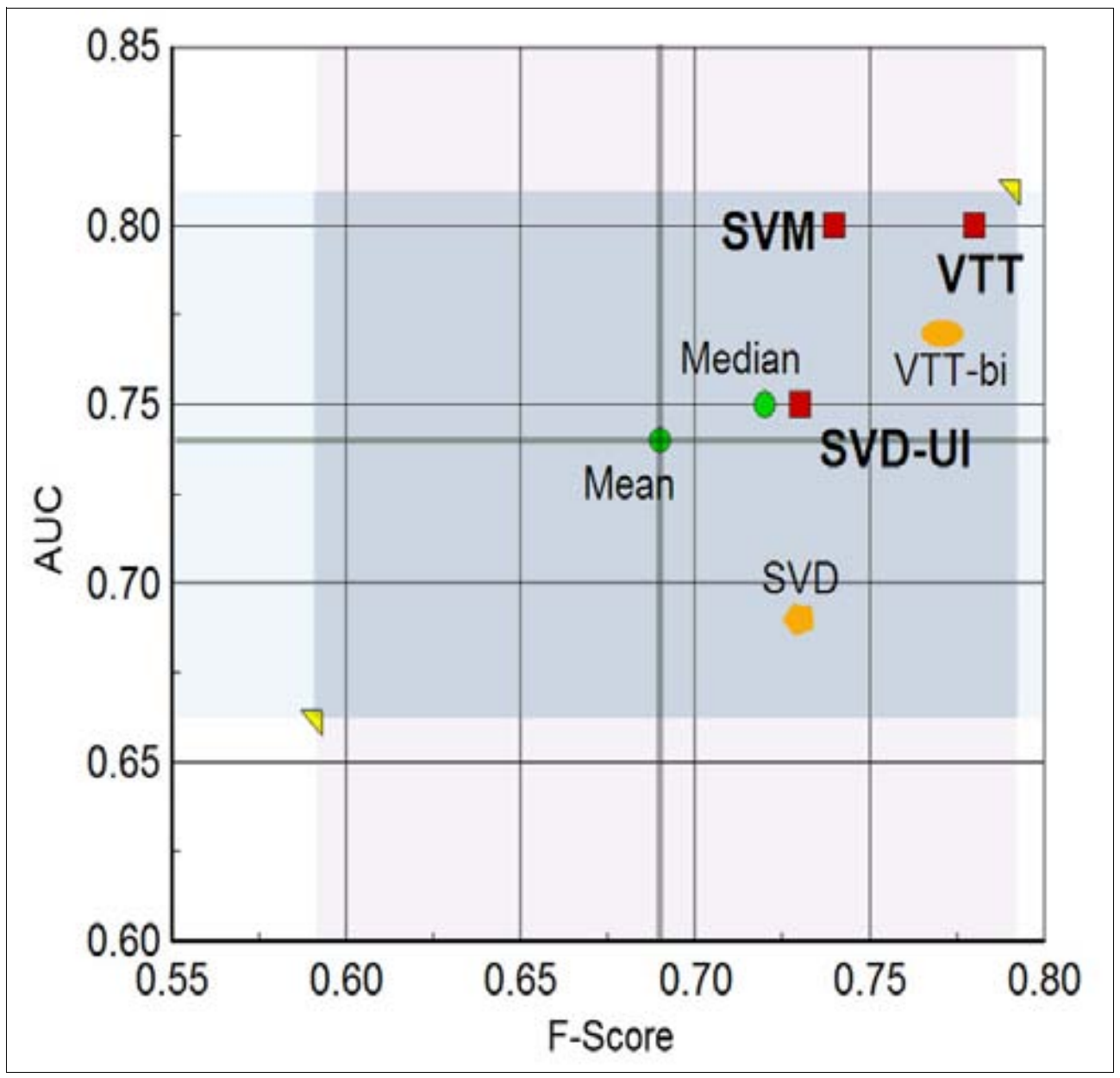

**Figure 3**
F-score versus AUC plane. Our methods on the F-score versus AUC plane for IAS. Mean and median are for the set of all submissions from all groups. Red squares denote our three submissions (SVM, VTT, and SVD-UI). The orange polygon denotes the results for SVD alone, and the orange oval denotes the results for one of the versions of VTT (with bigrams+) included in the SVD-UI method. AUC, area under the receiver operating characteristic curve; IAS, interaction article subtask; SVD, singular value decomposition; SVM, support vector machine; SVD-UI, SVD with uncertainty integration; VTT, variable trigonometric threshold.

Training our SVM classifier (see Materials and methods [below]) on the bootstrapped training set (rather than the original) resulted in an improved performance (AUC 0.82, accuracy 0.71, and F-score 0.75). This was the best AUC that we obtained, although the accuracy and F-score of the VTT method (0.75 and 0.78, respectively), trained on regular and not bootstrapped data, were still superior. This highlights our observation that our novel VTT method seems to suffer less from over-fitting to training data than both the SVM and SVD methods. Indeed, the performance of both the SVM and SVD on the training data was superior to that of the VTT. The mean values of F-score and accuracy on the eight training and eight additional data partitions (see Materials and methods [below]) for the SVD and SVM were around 0.92 (the SVM achieved 0.96 AUC on these training and additional data). In contrast, the VTT achieved values only around 0.88 for these same measures. Because the VTT surpassed both the SVM and SVD on the test data, we conclude that it generalized the 'concept' of protein interaction better.

It is important to note that our inclusion of additional data to train our algorithms (in particular the VTT) was very useful in achieving this generalization. Indeed, if we had trained exclusively on the supplied training data, then the VTT parameters would have been different: $\lambda_0 = 1.25$ and $\beta = 13$ (see Materials and methods [below]). If we had submitted this run, then the accuracy would have been the same as our submission (0.75), but both F-score and AUC would have been considerably worse, at 0.75 (from 0.78) and 0.75 (from 0.8), respectively. Thus, we conclude that the inclusion of additional data was useful in achieving a generalization of the 'concept' of protein interaction in the bibliome. Figure 4 depicts the decision surface for the VTT on the test data, as well as the decision surface that would have been submitted if we had trained exclusively on the training data supplied. Figures 5 and 6 depict the same surfaces but on one of the training and additional data partitions, respectively.

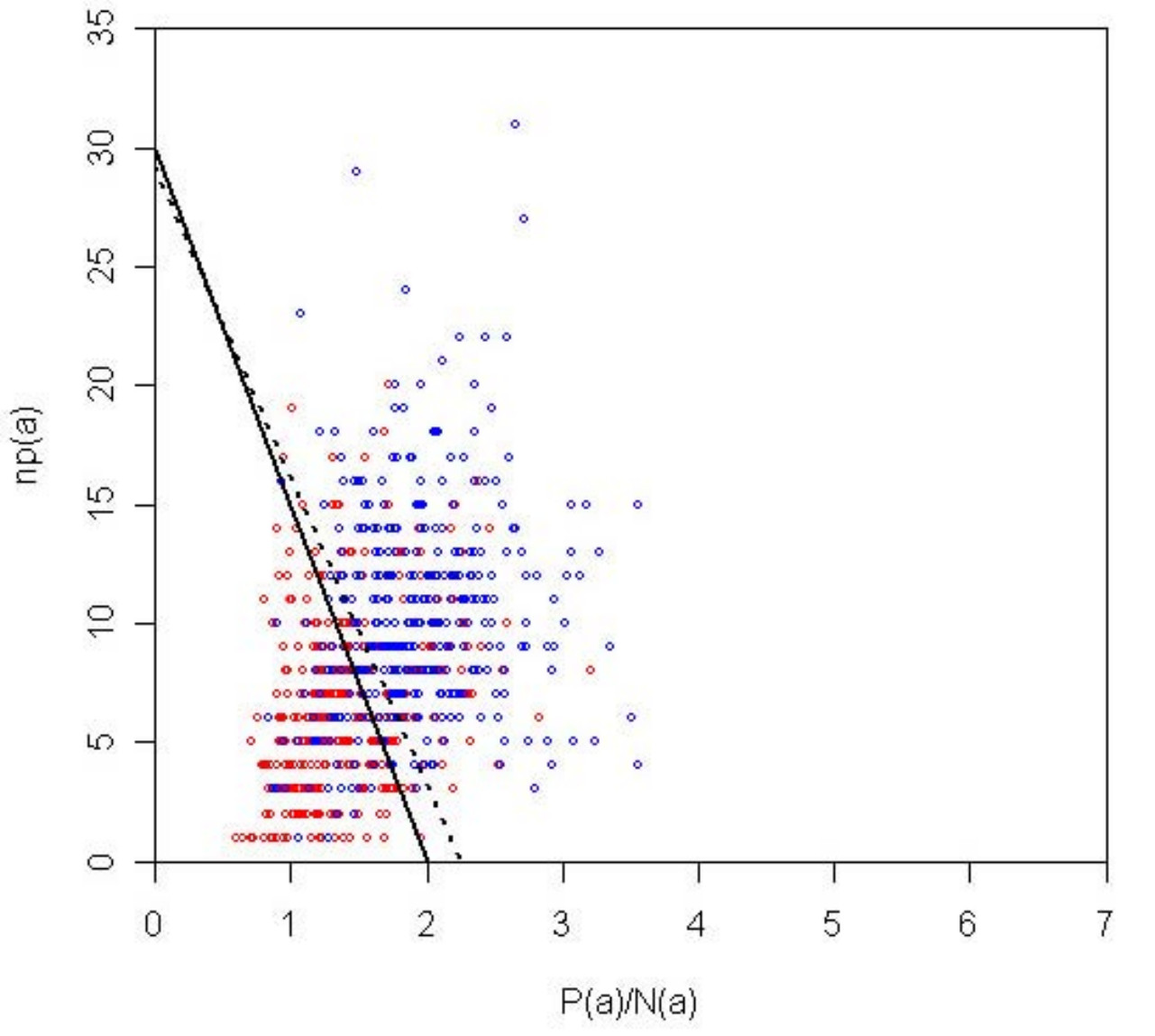

**Figure 4**
VTT decision surface for test data. Decision boundary for VTT on the space of *P*(*a*)/*N*(*a*) and *np*(*a*) for the test abstracts. Red and blue dots represent negative and positive abstracts. Dotted line represents surface that optimizes training data alone. VTT, variable trigonometric threshold.

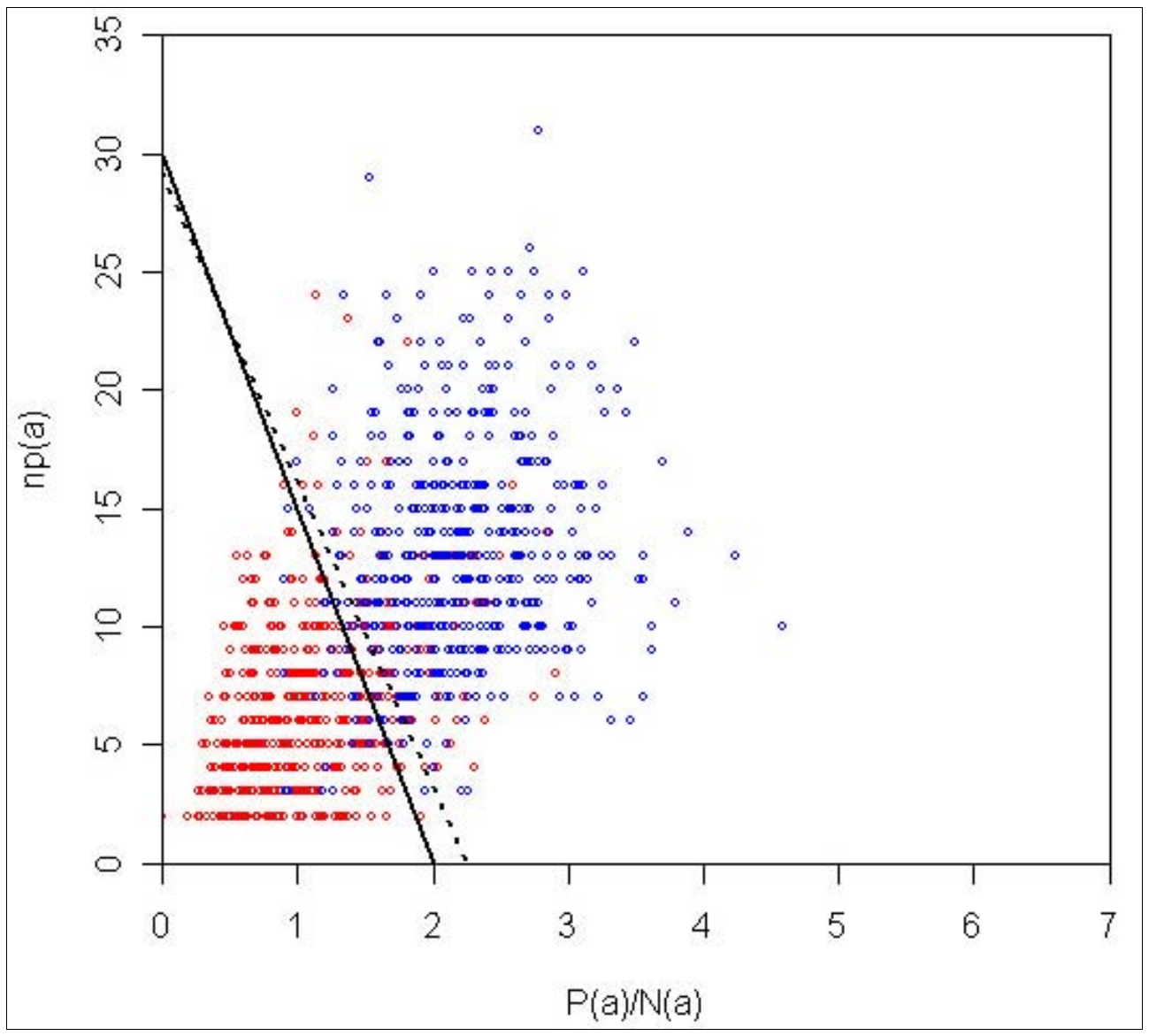

**Figure 5**
VTT decision surface for a training partition. Decision boundary for VTT on the space of *P*(*a*)/*N*(*a*) and *np*(*a*), for one of the training k-fold partitions. Red and blue dots represent negative and positive abstracts. Dotted line represents surface that optimizes training data alone. VTT, variable trigonometric threshold.

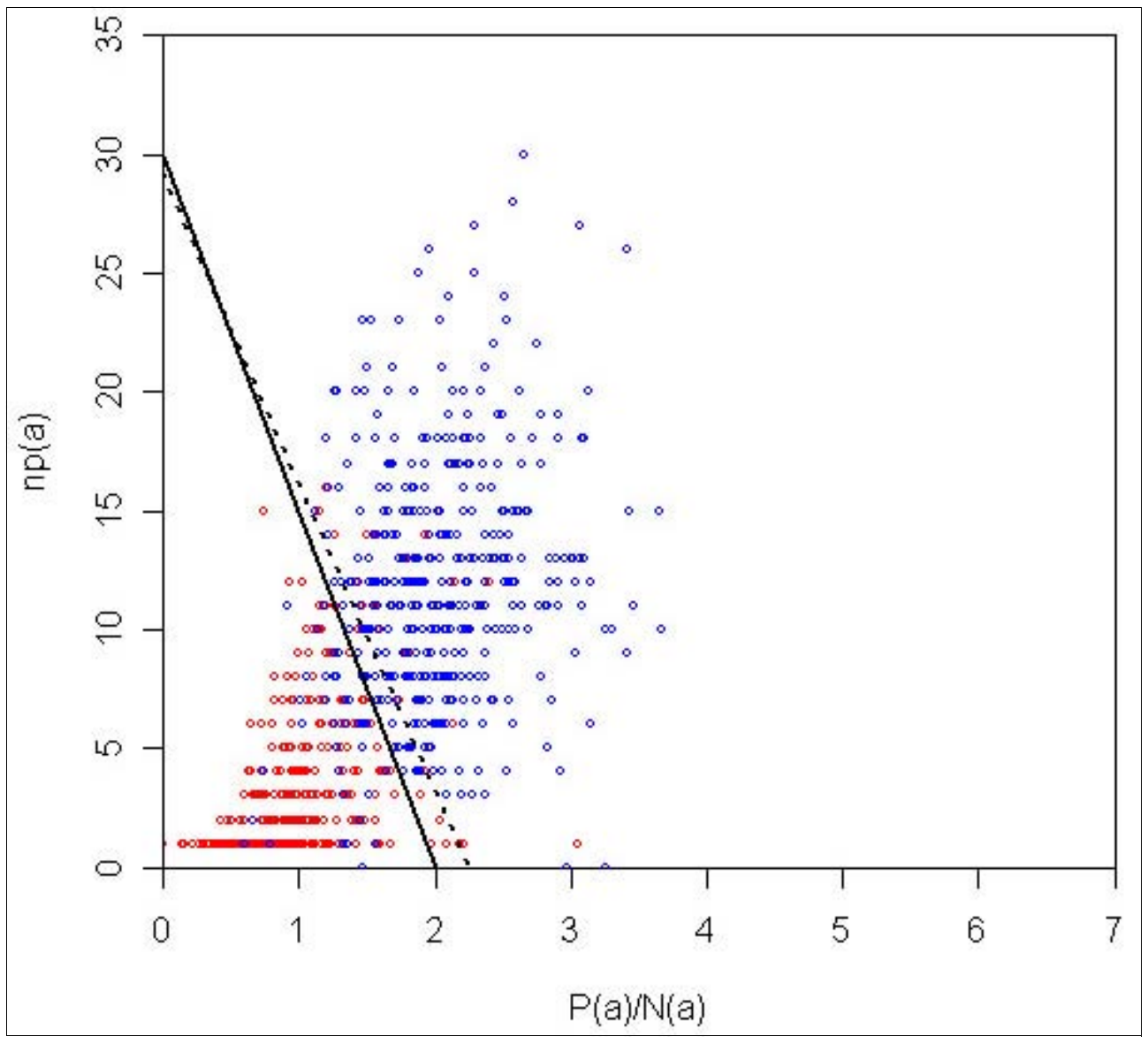

**Figure 6**
VTT decision surface for an additional data partition. Decision boundary for VTT on the space of *P*(*a*)/*N*(*a*) and *np*(*a*) for one of the additional data partitions. Red and blue dots represent negative and positive abstracts. Dotted line represents surface that optimizes training data alone. VTT, variable trigonometric threshold.

We should also note that our final VTT submission used abstract co-occurrence word pairs as features. The decision to submit a run with these features instead of bigrams+ was made based on the rank product of performance measures on training and additional data (as described in Materials and methods [below]). However, the performance of bigrams+ on training and additional data was just marginally inferior to co-occurrence pairs. When it came to the test data, however, bigrams+ led to a considerably lower AUC (0.77 from 0.8 obtained by co-occurrence pairs) and slightly lower F-score and accuracy. This seems to indicate that word co-occurrence word pairs in abstracts are also able to generalize relevance to protein interaction better than bigrams+. The bigram+ version of VTT (VTT-bi) was included in our uncertainty integration scheme and its performance can also be appreciated in Figures 1, 2, and 3.

### *PIARE tool*

We produced a web tool based on the VTT method, in order to classify abstracts according to protein interaction relevance: PIARE. The prototype of this tool is available on the internet [12]. A final version and web service is forthcoming. The PIARE tool receives an abstract (pasted or via a PMID) and issues a binary classification (relevant or nonrelevant to protein interaction) as well as a confidence factor provided by Equation 3 (presented in Materials and methods [below]). The confidence factor $C$ depends on the number of proteins identified in the abstract $np(a)$, and constants $\lambda_0$ and $\beta$. Currently, the tool arbitrarily assumes that if $C \leq 0.1$, then the confidence is considered low; if it is >50% ($C \geq 0.5$) then it is high; otherwise it is medium. In future work we intend to develop a more statistically sound measure of confidence based on random tests and user feedback.

## Interaction pair subtask and interaction sentences subtask results

The results for the IPS and ISS tasks, except for the recall measure of performance in IPS and mean reciprocal rank in ISS, were somewhat disappointing, although in line with the central tendency of all submissions. Our three submitted runs to IPS were barely distinguishable (within 1%). Runs 1 and 2 yielded the exact same results, which shows that using either the features extracted from IAS abstracts or those extracted from the IPS/ISS training sentences leads to the same results (see Materials and methods [below] for feature extraction details). On the other hand, run 3 yielded an improvement of about 1% over the two other runs, which shows that using both types of features simultaneously is somewhat advantageous. For all of our three runs, the precision was below the mean and median of all submissions but still well within one standard deviation of the mean. On the other hand, recall was above the mean and median of all submissions; very close to being above the mean plus one standard deviation for all articles; and above it for the subset of articles containing exclusively SwissProt IDs. There were 6 submissions (from 4 groups) out of 45 with higher recall for the set of all articles, and 7 for the case of articles with SwissProt IDs only (see [11] for more details). The F-score was very near to the mean and median of all submissions. Table 2 lists the details.

**Table 2**

**IPS results: summary of results of IPS task, including our three submissions and the central tendency values for all submissions.**

| | PPI all | | | PPI SP | | |
|---|---|---|---|---|---|---|
| | Precision | Recall | F-score | Precision | Recall | F-score |
| Run 1 | | | | | | |
| Run 2 | 0.051 | 0.275 | 0.072 | 0.056 | 0.285 | 0.077 |
| Run 3 | 0.052 | 0.278 | 0.073 | 0.057 | 0.288 | 0.078 |
| Mean[a] | 0.116 | 0.2 | 0.113 | 0.106 | 0.186 | 0.113 |
| StDev[a] | 0.104 | 0.106 | 0.084 | 0.095 | 0.1 | 0.084 |
| Median[a] | 0.081 | 0.216 | 0.084 | 0.076 | 0.196 | 0.084 |

Results are shown for the identification of protein-protein interaction pairs (PPI) for the set of all articles (All) and the subset of articles containing exclusively SwissProt interaction pairs (SP). [a]Calculated from 45 runs submitted by participating teams.

Regarding the ISS subtask, the three submitted runs were slightly different, and denoted a slight improvement with the number of the run. Run 2 was better than run 1, which shows that the proximity network expansion improved a little the original features - about 1% in terms of correct passages. Run 3 was better than run 2, showing that considering the paragraph rank from IPS (which includes number of protein mentions) in addition to the expanded word-pair features is advantageous - gaining another 1%. Our matches (387) and unique matches (156) to previously selected passages were above the average of all submissions (207.46 and 128.62, respectively) - there were only 3 out of 26 submissions (from 3 groups) with higher number of matches, and 8 (5 groups) out of 26 with higher number of unique matches [13]. One should note, however, that we predicted many more passages (18,371) and unique passages (5,252) than the average (6,213.54 and 3,429.65, respectively), which led to lower than average fractions of correct from predicted and unique passages. As in the case of IPS, this means that our system was better at recall than at precision. Finally, our mean reciprocal rank of correct passages was substantially higher than average (0.66 to 0.56); only 3 (from 1 group) out of 26 submissions were higher [13]. This shows that our system ordered correct passages better than most teams. Table 3 lists all of the values.

**Table 3**

**ISS results**

| | Average[a] | Run 1 | Run 2 | Run 3 |
|---|---|---|---|---|
| Predicted passages | 6,213.54 | 18,385 | 18,371 | 18,371 |
| Predicted unique passages | 3,429.65 | 5,156 | 5,270 | 5,252 |
| Matches | 207.46 | 360 | 376 | 387 |
| Unique matches | 128.61 | 131 | 145 | 156 |
| Fraction correct | 0.047 | 0.02 | 0.021 | 0.021 |
| Fraction unique correct | 0.047 | 0.025 | 0.028 | 0.03 |
| Mean reciprocal rank | 0.557 | 0.659 | 0.625 | 0.642 |

[a]Calculated from 26 runs submitted by participating teams. ISS, interaction sentences subtask.

## Conclusion

For IAS, we set out to test a novel lightweight classification method for protein interaction relevance, by contrasting it with two of the known best-performing classification methods: SVM and SVD. The lightweight VTT method uses a small number of original features (650) and a linear decision surface, but it outperformed our runs with SVM and SVD, as measured by accuracy, F-score and AUC, when using the same set of features. The VTT method also performed quite well when compared with all submissions to the BioCreative II challenge. The submissions from other groups that surpassed or were similar to VTT all used SVMs, but with more sophisticated features including semantic features from relevant biological ontologies and nomenclatures. In contrast, the VTT uses only the off-the-shelf named entity recognition package ABNER (A Biomedical Named Entity Recognizer) [14], and single word features selected by computing simple probabilities of occurrence in positive and negative training abstracts. In summary, our novel VTT method was good at generalizing the 'concept' of protein interaction from the bibliome, while remaining lightweight computationally by using only 650 word features. Training with additional data, using our rank product integration, was also shown to be advantageous.

We also tested an uncertainty-based integration method that we had previously developed for protein family prediction [4]. This integration did not improve the performance of the VTT, although it did improve the performance of the SVD.

Regarding the IPS task, although obtaining a good recall measure, our system could improve precision by considering additional biologically relevant information. In particular, because our system, for the same protein mention, outputs different Uniprot IDs for each of the organism MeSH terms of the document at stake, it could be improved by identifying organism information in the text. Using a classifier (such as our VTT or an SVM) to preclassify documents and passages according to different organisms could result in increased precision. We should also do more about removing genetic interactions. These were not considered in BioCreative, but we did not remove gene mentions from ABNER as features in our system. These and other improvements will be tested in the future.

Regarding the ISS task, our system was good at recalling and ranking predicted passages, but the feature expansion method based on word proximity networks could improve its precision. We used solely protein names to extract additional features from proximity networks; unfortunately, protein names as recognized by ABNER were often broken up by stemming and preprocessing of documents, appearing in different forms in the proximity networks and thus failing to be recognized by the feature expansion method. Because, even so, the expansion of feature words was modestly beneficial, and because it is clear from the manual observation of proximity networks that they do capture the contextual relationships of individual documents, we plan to use the method to find additional words related to general features, not just protein names. We also plan to use these document-specific networks to predict and classify documents according to organism, which should help improve protein pair prediction for tasks such as IPS and ISS.

In general, our participation in three subtasks of the BioCreative II challenge, with such a large set of members, was very useful in validating our approaches as well as learning from other groups. It also led us to a position where we are more easily able to extend the methods to biomedical applications other than protein interaction.

## Materials and methods

### Protein interaction article subtask

As mentioned under Background (above), most text-mining applications in biomedicine start with the collection of relevant documents, typically from abstract information. In the case of the IAS, the goal is to classify 'articles which contain relevant information relative to protein interactions' [6]. The groups participating in BioCreative II were first given training data consisting of a collection of PubMed article abstracts that were previously curated according to IntAct [7] and MINT [8] standards. Note that such curation used full-text articles, whereas training data contained only abstracts. Thus, this subtask assumes that it is in general possible to decide, solely from an article's abstract, whether it contains protein interaction information in its full-text content. The training data contained the following: 3,536 abstracts categorized as PPI relevant, known as true positives (TPs); 1,959 irrelevant abstracts known as true negatives (TNs); and an additional set of about 13,000 abstracts considered likely TPs, extracted from alternative interaction databases. The unlabeled test data consisted of 750 abstracts, and was given to participant groups only 1 week before the submission deadline. We submitted three runs using the different methods outlined below.

#### *Feature selection*

All three runs submitted use 650 word features extracted from the training data using a method broadly based on the spam filtering system SpamHunting [15]. First, we computed the probability $p_{TP}(w)$ that a word $w$ appears in a positive abstract, as the ratio of the number of positive abstracts containing $w$, over the total number of positive abstracts. Similarly, we computed the probability $p_{TN}(w)$ that a word $w$ appears in a negative abstract, as the ratio of the number of negative abstracts containing $w$, over the total number of negative abstracts. After stemming with the Porter algorithm, filtering out short words with two or fewer letters, and removing common stop words except the word 'with', we ranked all words according to the score: $S(w) = |p_{TP}(w) - p_{TN}(w)|$. (The word 'with' was observed to appear in many bigram and word pairs associated with positive abstracts [as described below] in tests with no stop word removal; therefore, we kept it because it appears to be important in the syntax of PPI expressions.) The words with the highest score $S$ tend to be associated either with positive or negative abstracts. Therefore, such words are assumed to be good features for classification. In order to keep the system very lightweight, we used only the top 650 stemmed abstract words with largest $S$ as our continuent features; the top 15 words are listed in Table 4. We selected the top 650 stemmed words because this is the value at which the histogram of $S$ per ranked words becomes essentially flat (null derivative). Furthermore, more aggressive pruning of features (fewer features) resulted in worse performance on test training data (as described below), and more conservative pruning of features (more features) resulted in no improvement.

We produced two additional feature sets made up of word pairs obtained from the 650 stemmed word features in the first set. This leads to $650^2$ = 422,500 possible word pairs, although not all occur. First, we removed all words not in the first feature set from the abstracts. Then, from these filtered abstracts (vectors), we obtained the second and third feature sets, which are comprised of 66,301 pairs of words immediately adjacent in the filtered abstract vectors (bigrams$^+$) and 48,006 unique pairs of words that co-occur in these vectors, respectively. We also computed the probability that such word pairs $(w_i, w_j)$ appear in a positive or negative abstract: $p_{TP}(w_i, w_j)$ and $p_{TN}(w_i, w_j)$, respectively. Figure 7 depicts the 1,000 abstract co-occurrence word pairs (the third feature set) with largest $S^{ab}(w_i, w_j) = |p_{TP}(w_i, w_j) - p_{TN}(w_i, w_j)|$, plotted on a plane where the horizontal axis is the value of $p_{TP}(w_i, w_j)$ and the vertical axis is the value of $p_{TN}(w_i, w_j)$; we refer to this as the $p_{TP}/p_{TN}$ plane. Table 4 lists the top 15 word pairs for $S^{ab}$. Plots and additional data for all feature sets are included in supplemental materials that are available online [16].

**Table 4**

**Top 15 word and word pair features for IAS task ranked by $S$ and $S^{ab}$ measures.**

| Top 15 words for $S$ | | | | Top 15 pairs for $S^{ab}$ | | | |
|---|---|---|---|---|---|---|---|
| $w$ | $p_{TP}$ | $p_{TN}$ | $S$ | $w_i$, $w_j$ | $p_{TP}$ | $p_{TN}$ | $S^{10}$ |
| interact | 0.76 | 0.12 | 0.64 | with, interact | 0.56 | 0.06 | 0.50 |
| bind | 0.63 | 0.14 | 0.49 | interact, protein | 0.23 | 0.02 | 0.21 |
| domain | 0.52 | 0.08 | 0.44 | between, interact | 0.25 | 0.12 | 0.13 |
| complex | 0.46 | 0.15 | 0.31 | two-hybrid, yeast | 0.18 | 0.00 | 0.17 |
| proteom | 0.01 | 0.29 | 0.28 | associ, with | 0.27 | 0.10 | 0.16 |
| with | 0.9 | 0.65 | 0.25 | bind, protein | 0.19 | 0.03 | 0.15 |
| yeast | 0.28 | 0.04 | 0.24 | with, complex | 0.16 | 0.02 | 0.14 |
| activ | 0.55 | 0.32 | 0.23 | analysi, proteom | 0.00 | 0.12 | 0.12 |
| two-hybrid | 0.23 | 0.00 | 0.22 | with, protein | 0.24 | 0.13 | 0.11 |
| protein | 0.86 | 0.64 | 0.22 | bind, domain | 0.12 | 0.02 | 0.1 |
| between | 0.38 | 0.16 | 0.22 | interact, domain | 0.10 | 0.00 | 0.1 |
| associ | 0.35 | 0.13 | 0.22 | protein, domain | 0.10 | 0.01 | 0.09 |
| region | 0.26 | 0.06 | 0.2 | with, domain | 0.1 | 0.01 | 0.08 |
| function | 0.48 | 0.28 | 0.2 | mass, spectrometri | 0.02 | 0.10 | 0.08 |
| regul | 0.38 | 0.19 | 0.19 | two-hybrid, screen | 0.08 | 0.00 | 0.08 |

IAS, interaction article subtask.

One should note that our bigrams$^+$ are built only from the 650 single word features, and therefore they are not necessarily constituted of words immediately adjacent in abstracts. They include traditional bigrams only if both words are in the set of 650 single word features. However, they also include pairs of words that are not necessarily adjacent in an abstract, but are adjacent in the word vectors comprised of only the top 650 single word features produced for each abstract. As for the abstract co-occurrence word pairs, all of these co-occur in the same abstracts, but they are likewise comprised of only the 650 single word features.

In addition to the top 650 word, bigram$^+$, and abstract word co-occurrence features, we used the number of unique protein mentions per abstract $a$, $np(a)$, as an additional feature or parameter. To compute this number, we used ABNER [14,17].

*Training and additional data*

To train the various classification methods described below, we first performed k-fold tests on the supplied training data. Specifically, we randomly generated eight different partitions of the training set of abstracts, with 75% of the abstracts used to train the classification algorithms employed, and 25% to test them. In addition, we forced the 25% test sets of abstracts in these partitions to have a balanced number of positive (TP) and (TN) negative abstracts. We conducted a second test using additional data not supplied by the BioCreative II organizers. We collected 367 additional positive abstracts from the MIPS (Munich Information Center for Protein Sequences) database [18], and 427 negative proteomics abstracts curated by hand that were graciously donated to our team by Santiago Schnell. The second test then consisted of training the classification algorithms with all of the supplied positive and negative abstracts (TP and TN), and testing on the additional data that were also balanced with the addition of 60 randomly selected, likely positive abstracts from TP. We produced eight different randomly selected balanced test sets with the additional data. Finally, we used the k-fold and additional data tests to select the best parameters for the various classification algorithms employed, as described below.

*Variable trigonometric threshold classification*

It is obvious that the best feature terms in the $p_{TP}/p_{TN}$ plane are the ones closest to either one of the axes. (By 'term', we refer to features in our three different feature sets.) Any feature term $w$ is a vector on this plane (see Figure 8), and therefore term relevance to each of the classes can be measured with the traditional trigonometric measures of the angle $\alpha$, of this vector with the $p_{TP}$ axis: $\cos(\alpha)$ is a measure of how strongly terms are exclusively associated with positive abstracts, and $\sin(\alpha)$ with negative ones (in the training data). Then, for every abstract $a$, we compute the sum of all feature term contributions for a positive (P) and negative (N) decision:

$$P(a) = \sum_{w \in a} \cos(\alpha(w)), \quad N(a) = \sum_{w \in a} \sin(\alpha(w)) \qquad (1)$$

Given the nature of the BioCreative 2 challenge, with a short time to submit results, we did not analyze the distributions of feature contributions using the trigonometric measures above. We will leave that analysis for future work in an

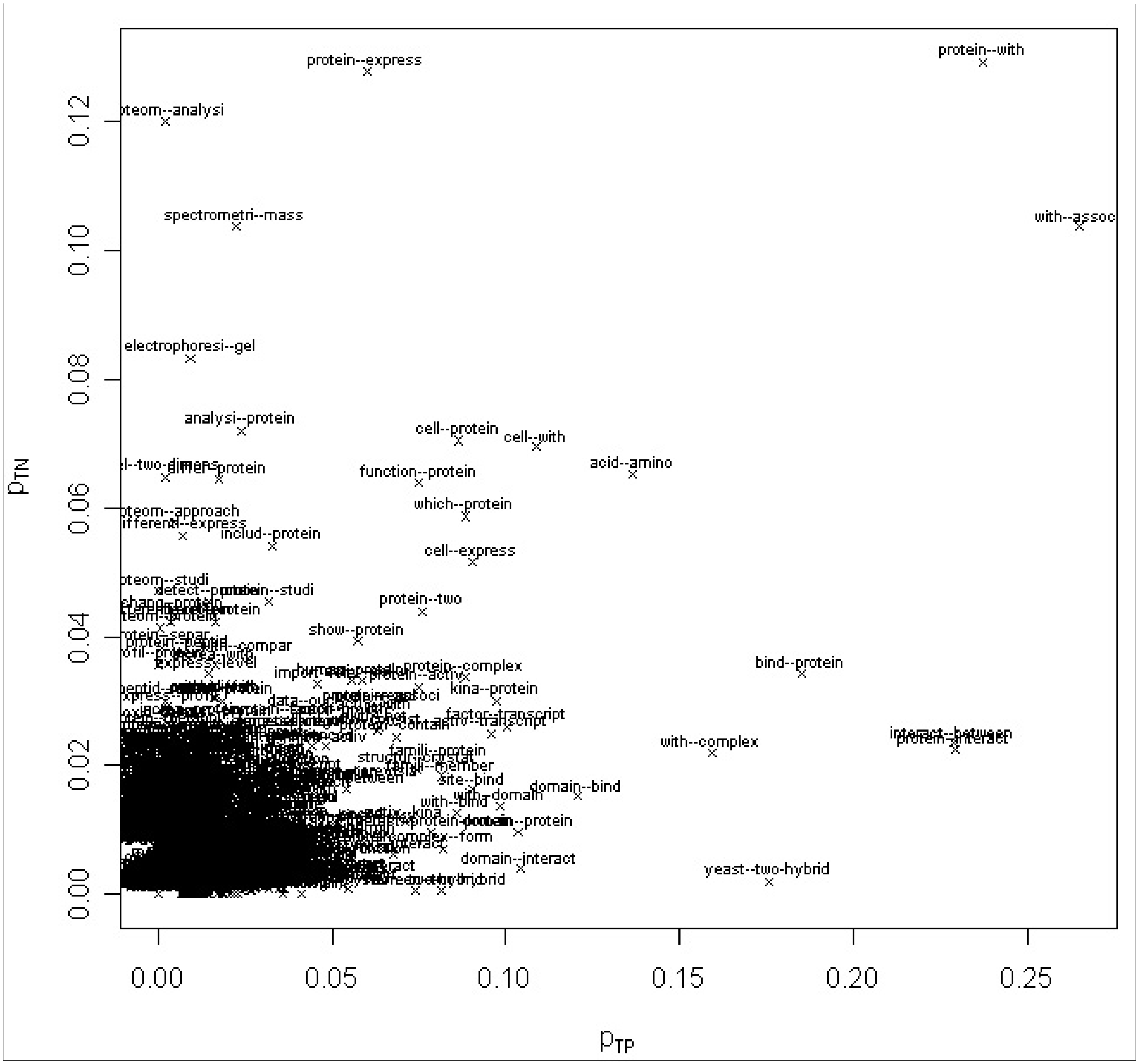

**Figure 7**

Top 1,000 abstract co-occurrence word pairs in the space of $P_{TP}(w_i, w_j)$ and $P_{TN}(w_i, w_j)$. Word pairs with high $S^{ab}$ tend to be almost exclusively associated with positive or negative abstracts.

attempt to better justify theoretically our feature weight aggregation. However, we can report that using our trigonometric measures led to better performance than summing $p_{TP}$ and $p_{TN}$ alone in our tests (described below).

The decision of whether abstract $a$ is a positive or negative abstract (insofar as being relevant to protein-protein interaction) is then computed as follows:

$$\begin{cases} a \in TP, & \text{if } \frac{P(a)}{N(a)} \geq \lambda_0 + \frac{\beta - np(a)}{\beta} \\ a \in TN, & \text{otherwise} \end{cases} \quad (2)$$

Where $\lambda_0$ is a constant threshold for deciding whether an abstract is positive (relevant) or negative (irrelevant). This threshold is subsequently adjusted for each abstract $a$ with the factor $(\beta - np(a))/\beta$, where $\beta$ is another constant, and $np(a)$ is the number of protein mentions in abstract $a$ as described in the feature selection subsection. (An alternative interpretation of the decision criteria is when the factor $(\beta - np(a))/\beta$ is moved to the left side of the inequality 2; in this case, the threshold is constant, but the feature contributions are adjusted for each abstract according to the number of proteins $np(a)$ and constant $\beta$.) We observed, in the training data, that abstracts have a higher chance of being positive (relevant) with more protein mentions; thus, via formula 2, the classification threshold is linearly decreased as $np$ increases. This means that with a higher (lower) number of protein mentions, it is easier to classify an abstract as positive (negative). When $np(a) = \beta$ the threshold is simply $\lambda_0$. We refer to this classification method as VTT. Listing positive and negative abstracts separately, the ranking of abstracts is straightforward from formula 2: abstracts are ordered in decreasing value of $|P(a)/N(a) - T(a)|$, where $T(a) = \lambda_0 + (\beta - np(a))/\beta$ is the threshold point for abstract $a$. The further away from the decision surface an abstract is, the higher it is in the respective ranking. In fact, from this ranking we can derive a confidence factor $C$ of the decision made for every abstract $a$:

$$C = \frac{\left|\frac{P(a)}{N(a)} - T(a)\right|}{|T(a)|} \quad (3)$$

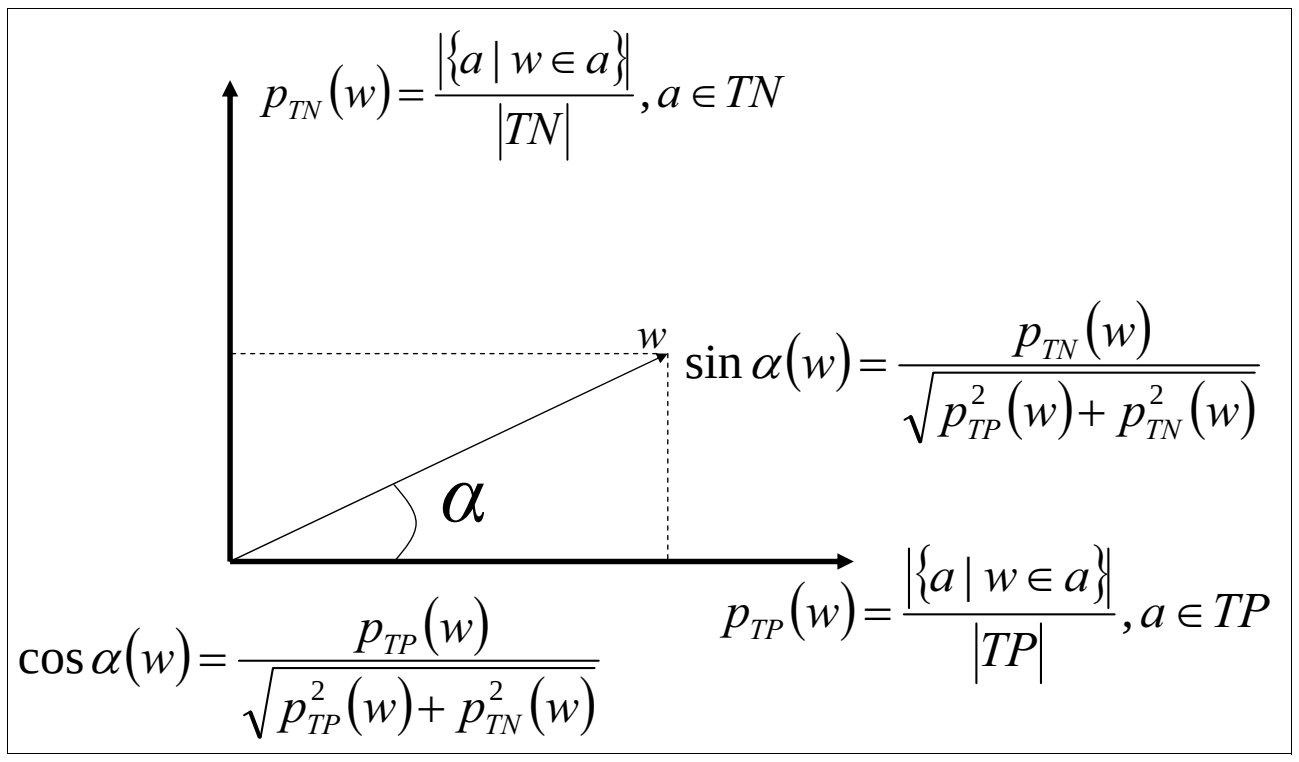

**Figure 8**
Trigonometric weights for features. Trigonometric measures of term relevance for identifying positive and negative abstracts in the $P_{TP}$ and $P_{TN}$ planes.

This confidence factor is a measure of the distance of an abstract's ratio of feature weights $(P(a)/N(a))$ to the decision surface, or threshold point for the abstract $T(a)$.

The final values of $\lambda_0$ and $\beta$ submitted were determined by optimizing classification performance on the training tests described above (k-fold and additional data tests). We swept the following range: $\lambda_0 \in [0,10]$ and $\beta \in [1,50]$, in intervals of $\Delta\lambda = 0.25$ and $\Delta\beta = 2$. For each $(\lambda_0, \beta)$ combination, we computed the mean and variance of F-score and accuracy measures for the eight k-fold tests and eight additional data tests. We computed four ranks for the classifiers tested in the parameter range: $r_F^K$ and $r_A^K$ rank classifiers according to the mean value of F-score and accuracy in the eight k-fold tests of the training data, respectively; and $r_F^T$ and $r_A^T$ rank classifiers according to the mean value of F measure and accuracy in the eight additional data tests, respectively. We then ranked all classifiers tested according to the rank product of these four ranks: $R = r_F^K \times r_A^K \times r_F^T \times r_A^T$ . This procedure was performed for the three distinct feature sets: 650 single words with largest $S$, bigrams$^+$, and abstract co-occurrence pairs. Finally, we submitted a run with VTT using the parameters that yielded the smallest value of $R$, which was 640. The parameters were $\lambda_0 = 1$ and $\beta = 15$, using the third feature set (abstract co-occurrence words). Thus, this combination optimized F-score and accuracy on the training and additional data. It is worth mentioning that the parameter set that optimizes these measures on both training and additional data is not the same one that optimizes results on training data alone ($\lambda_0 = 1.25$ and $\beta = 13$, also with the third feature set). Figures 5 and 6 depict the decision surface of the VTT model with the submitted parameters for one of the training partitions and one of the additional data partitions, respectively.

To compare the performance of the lightweight VTT system above, we tested it against machine learning methods that typically yield the best performance in classifications tasks: SVMs [19] and the SVD [20]. The next subsections detail our approach to these methods.

### *Support vector machine classification*

To best compare this method with VTT, we started from the same original feature set: the 650 single words with largest $S$. We applied additional dimensionality reduction and then trained classification models to discriminate between positive and negative data. Dimensionality reduction involved a two-step process. First, a feature selection filter based on the $t$-test was used, in which all features with the $p$ value below a pre-specified threshold $t_f$ were retained. Then, we applied the principal component analysis [20] to retain all features containing $t_{PCA} \times \sigma^2$ of the total variance $\sigma^2$. The remaining features were fed into a SVM, a classification model used to maximize the margin of separation between positive and negative examples [19]. We used the SVM$^{light}$ package [21], in which we explored both polynomial and Gaussian kernels with various parameters. The overall system was trained to maximize the classification accuracy on the unlabeled data using the following two-step iterative procedure: (i) train a classifier with costs adjusted to the current estimates of class priors in the unlabeled data; and (ii) predict class labels on the unlabeled set using current classifier and make new estimates of the class priors. Initially, class priors in the unlabeled data were set to 0.5. Not more than five rounds were executed, ending with the total cost of positive examples being about three times the costs of the negatives. The final predictor, which we submitted as run 1, used $t_f$ = 0.1 for the feature filtering, $t_{PCA}$ = 0.95 for the principal component analysis, and a linear SVM. One should note that we also tested the SVM classification starting from all stemmed words in the set of training abstracts, rather than just the 650 from the first feature set, followed by the same process of dimensionality reduction leading to at most 2,000 features. Testing our SVM with this feature selection method on the eight k-fold training data and eight additional data partitions (as well as on the test data itself after the challenge) yielded no gains in performance, suggesting that our selection of the top 650 words with largest $S$ for VTT is sufficient for classification.

*Singular value decomposition classification*

To best compare this method with VTT, we started from the same original feature set: the 650 single words with largest $S$. We represented abstracts as vectors in this feature space. We then calculated the inverse document frequency (IDF) measure, so the vector coefficients were the TF*IDF [22] for the respective features. The number of protein mentions per abstract, $np(a)$ (see Feature selection subsection), was added as an additional feature. The abstract vectors were also normalized to Euclidean length 1. We computed the SVD [20] of the resulting abstract-feature matrix (from the training data). The top 100 components were retained (this number provided best results on our tests on training and additional data).

We classified the set of abstracts using a nearest neighbor classifier on the eigenvector space (of dimension 100) obtained via the SVD of the feature/abstract matrix. To classify a test abstract vector $a$, we project it onto this SVD subspace and calculate the cosine similarity measure of $a$ to every training abstract $t$:

$$\cos(a,t) = \frac{a.t}{||a|| \times ||t||} \tag{4}$$

We then calculate positive and negative scores for each test abstract $a$ by summing the cosine measure for every positive ($t \in TP$) and negative ($t \in TN$) training abstract, respectively:

$$\pi(a) = \frac{1}{|TP|}\sum_{t \in TP} \cos(a.t), \quad \nu(a) = \frac{1}{|TN|}\sum_{t \in TN} \cos(a.t) \tag{5}$$

Where $|TP|$ and $|TN|$ are the number of positive and negative abstracts in the training data, respectively. (Often, the aggregation of vector contributions would be made for the nearest K vectors [or a neighboring hypercone in vector space] rather than summing the contributions of every vector $t$ in the space. Using all training vectors could result in distortions by the existence of large masses of vectors in an opposing class. However, in the case of our balanced training sets these distortions are very unlikely. Moreover, in our previous work [4], use of all vectors resulted in better results than nearest K or nearest hypercone.) Finally, a linear decision boundary was determined in the two-dimensional space of $\pi$ and $\nu$; abstract $a$ is classified as positive (relevant) if $\pi(a) > m \times \nu(a) + b$ and as negative otherwise. Coefficients $m$ and $b$ were determined manually from optimizing the F-score measure on the eight k-fold training data and eight additional data partitions. Figure 9 depicts the boundary surface in the $\pi$ and $\nu$ space for training and test abstracts. One should note that we also tested our SVD method using the word pair features (bigrams+ and abstract co-occurrence word pairs), but there was no performance improvement on the training data. We did not submit a run exclusively with SVD for the challenge, but we tested it and integrated it in the method described in the following subsection.

*Uncertainty integration classification*

Using a variation of a method we previously introduced for protein family prediction [4], we integrated two variations of the VTT classification method with the SVD classification in such a way that for each abstract the most 'reliable' prediction was used to issue a classification. To ascertain reliability, we represented the target test abstract $a$, as well as all abstracts $t$ in the training data, as vectors in a compound feature space (including all three feature sets). Next, we computed the cosine similarity, $cos(a, t)$, between a target $a$ and every $t$ (formula 4), and treated this value as a weighted vote. Thus, if abstract $t$ is very close to $a$, then it will have a greater influence in the classification of $a$. Because for any abstract $t$ in the training data, we know whether a given classification method correctly classified it, we can measure reliability using Shannon's entropy - as in our previous work [4]. We compute the uncertainty (entropy) of a prediction for the target abstract based on the distribution of positive and negative weighted votes obtained for that abstract from a given classification method. Let $\rho_M(a, TP)$ and $\rho_M(a, TN)$ denote the probabilities of predicting, using method $M$, that abstract $a$ is positive (TP) or negative (TN), respectively. We estimate these probabilities as follows:

$$\rho_M(a,TP) = \frac{\Sigma_{t \in TP} \, cos(a,t)}{\Sigma_{t \in TP \cup TN} \, cos(a,t)}, \quad \rho_M(a,TN) = \frac{\Sigma_{t \in TN} \, cos(a,t)}{\Sigma_{t \in TP \cup TN} \, cos(a,t)}$$

Note that $\rho_M(a, TP) = 1 - \rho_M(a, TN)$. Finally, we compute the prediction uncertainty of abstract $a$ using method $M$, $U_M$(a), with Shannon's entropy as follows:

$$U_M(a) = -\rho_M(a, TP) \log \rho_M(a, TP) - \rho_M(a, TN) \log \rho_M(a, TN)$$

Using this uncertainty measure we integrate the predictions issued by each method by selecting, for each abstract $a$, the prediction issued by the method $M$ with lowest $U_M$(a); this value of uncertainty is also used to rank the abstracts for relevance. In our original submission to the BioCreative II chal-lenge, we submitted a run (run 3) based on this uncertainty-driven integration method with additional characteristics described in detail in [11]. Here, we report on updated results (run 3') after fixing the software error that afflicted the original VTT submission (run 2). Specifically, our SVD-UI scheme integrated three methods.

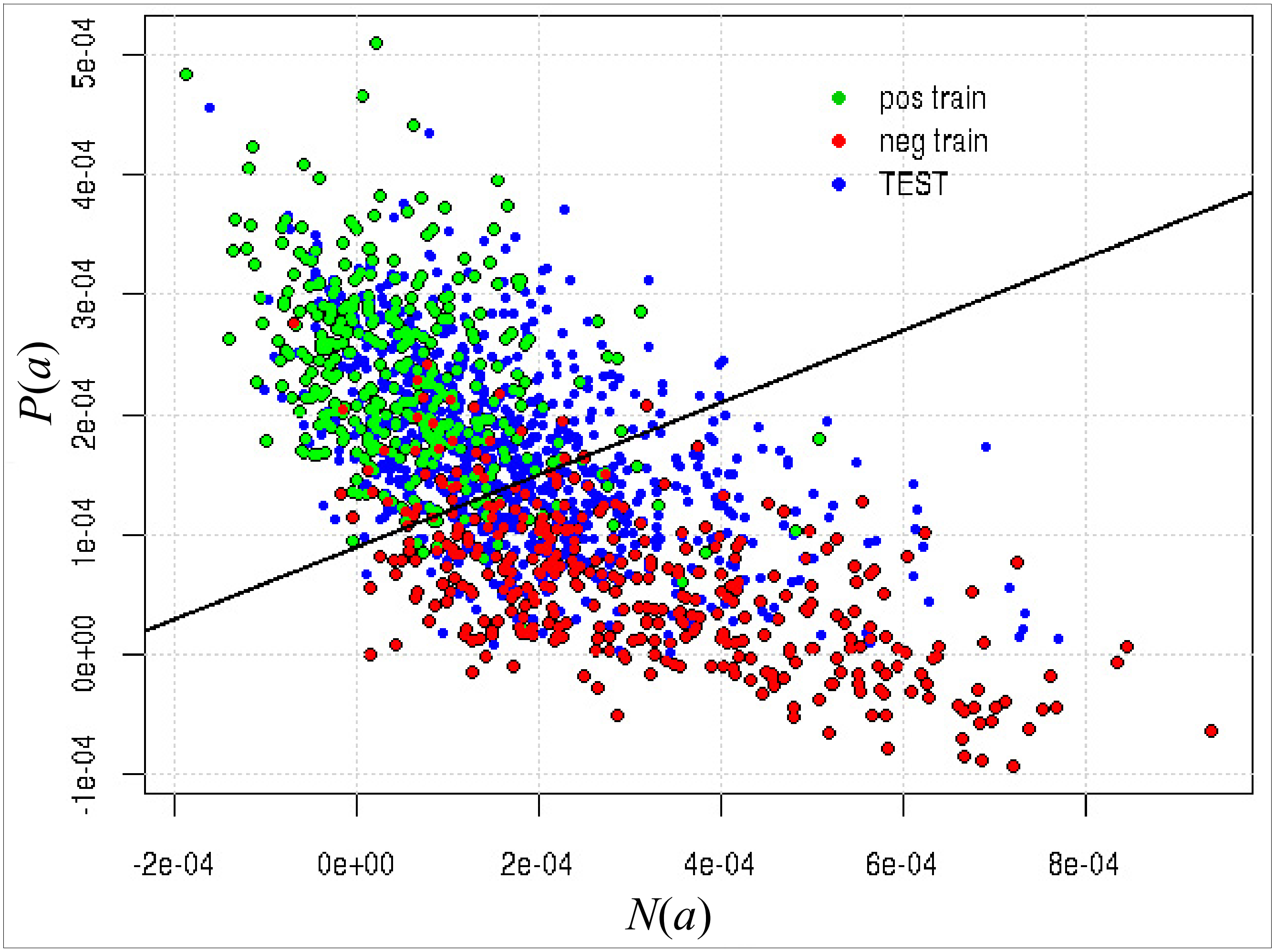

**Figure 9**
SVD decision surface for training and test data. Decision boundary for vector/SVD model on the space of positive and negative scores $\pi$ and $\nu$. Red and green dots represent negative and positive abstracts in the learning data set, respectively. Blue dots represent unlabeled test set abstracts. SVD, singular value decomposition.

1. SVD vector model with first feature set of single words.

2. VTT exactly as submitted in Run 2' (described above), using the third feature set (abstract co-occurrence word pairs) with $\lambda_0 = 1$ and $\beta = 15$.

3. VTT using the second feature set (bigrams$^+$) with $\lambda_0 = 1.5$ and $\beta = 9$. These parameters led to the best results using the second feature set, after our tests on the eight training and eight additional data partitions ($R = 860$).

Items 2 and 3 were chosen so that there would be a model from each of the word pair feature sets. It is important to note that in our tests with training and additional data, the SVD-UI improved only very slightly over the SVD vector model alone. Indeed, for the test set the SVD vector model alone produced the same relevant/nonrelevant classification as the integration method; the difference was only in the ranking of abstracts, thus affecting only the AUC performance measure, as discussed in Results (above). This was true for both the run submitted to the challenge (run 3) and the updated version (run 3'), as shown in Table 1.

The fact that SVD-UI and SVD alone yielded the same relevant/nonrelevant classification, indicates that when abstracts are projected onto the compound vector space described above, the classification via SVD is less uncertain (lower Shannon entropy) than the one via VTT. By this we mean that abstracts deemed positive (negative) by SVD tend to have less negative (positive) abstracts around them in the compound vector space (as measured by cosine similarity) than those classified by VTT. We decided to submit the results of the SVD-UI method other than SVD on its own, because it led to slightly better AUC measure results than the SVD vector model on the learning and additional data (see Results [above]). Thus, although SVD and SVD-UI classified the abstracts in the same manner, they led to different rankings. This indicates that using Shannon's measure of entropy on the compound vector space yields a better ranking than distance from the SVD decision surface alone.

## Protein interaction pair and sentences subtasks

Whereas the IAS subtask in the BioCreative II challenge focused on detecting relevant articles for extracting protein interaction information from collections of abstracts, the IPS and ISS focused on discovering and extracting the actual protein interaction information from individual, full-text documents. IPS focused on identifying pairs of proteins, described by their UniProt IDs [23], whose interaction is reported in a given article, and ISS on extracting the passages (containing at most three sentences) where such interaction is described and reported. For these two subtasks, the groups participating in BioCreative II were given training data consisting of 740 and a test set of 358 full-text articles [13]. For IPS training purposes, annotation files, containing normalized interaction pairs, for each article in the training set were also provided. Some sentence training data was also provided for ISS: 63 evidence passages from the training collection of articles, plus some additional sets of sentences derived from other resources (for details, see [13]). We approached IPS and ISS essentially as if they were a single task, and submitted three runs for each using the different methods outlined below.

### *Feature selection*

From the features extracted from abstracts in the IAS subtask, we collected 1,000 abstract co-occurrence word-pair features, $(w_i, w_j)$, from the third feature set. Because the purpose of these tasks is to identify portions of text in which PPI information appears, we do not need to worry about features indicative of negative PPI information. Thus, these features were chosen and ranked according to the highest values of the following:

$$p(w_i, w_j) = p_{TP}(w_i, w_j) \cdot cos(\alpha(w_i, w_j)) = \frac{p_{TP}^2(w_i, w_j)}{\sqrt{p_{TP}^2(w_i, w_j) + p_{TN}^2(w_i, w_j)}} \quad (6)$$

Where $p_{TP}$ and $p_{TN}$ are as defined in the IAS task methods subsection. This measure is a variation of the trigonometric measures we used in the VTT model for the IAS subtask. We multiply the cosine measure by the probability of the feature being associated with a positive abstract, to ensure that the many features which have zero probability of being associated with a negative abstract ($P_{TN} = 0$) are not equally ranked. Using Equation 6, when $P_{TN} = 0$, $p(w_i, w_j) = p_{TP}(w_i, w_j)$, rather than the 1 we would obtain if we did not use the additional $p_{TP}(w_i, w_j)$ factor. We refer to this set of 1000 stemmed word pairs, as the 'word pair feature set'.

We also obtained an additional set of features from PPI-relevant sentences: the 'sentence feature set'. These sentences were extracted from all PPI evidence sentences provided by BioCreative II for these tasks; these contained the 63 sentences associated with the set of training articles, as well as the sentences extracted from other resources detailed in [13]. From these PPI evidence sentences, we calculated the frequency of stemmed words: $f_{ppi}(w)$. Then, we calculated the frequency of stemmed words of the entire training corpus of 740 full-text articles: $f_c(w)$. Finally, similarly to the word pair features above, we selected as sentence features the top 200 stemmed words which maximize the following score (top 10 in Table 5):

**Table 5**

**Top 10 protein-interaction evidence sentence features used in IPS and ISS.**

| Rank | Feature |
|---|---|
| 1 | with |
| 2 | protein |
| 3 | cell |
| 4 | interact |
| 5 | bind |
| 6 | activ |
| 7 | express |
| 8 | complex |
| 9 | dna |
| 10 | human |

IPS, interaction pair subtask; ISS, interaction sentences subtask.

$$SS = \frac{f_{ppi}^2(w)}{\sqrt{f_{ppi}^2(w) + f_c^2(w)}} \quad (7)$$

### *Paragraph selection and ranking*

Our next step was to select paragraphs in each document that are more likely to contain protein interaction information. For this we used our two feature sets defined in the previous subsection, plus protein mention information. Thus, for each full-text document, we ordered paragraphs according to three different preference criteria.

A Largest sum of word pair feature weights, where the weights are the inverse feature rank. Paragraphs without feature matches are thrown out (rank 0).

B Largest number of protein mentions in paragraph. As in the IAS subtask, we also used ABNER to collect protein mentions in the full-text documents provided for these two subtasks.

Paragraphs without protein mentions are thrown out (rank 0).

C Largest number of sentence features in the paragraph. Each feature that occurs in a paragraph adds 1 to the count. Paragraphs without feature matches are thrown out (rank 0).

From these three distinct paragraph preference orderings, for each document we produced another three rankings that aim to integrate this information in different ways. For each document, we rank paragraphs according to the following criteria.

1. Rank product of preference orderings from A (word pair features) and B (protein mentions) above.

2. Rank product of preference orderings from B (protein mentions) and C (sentence features) above.

3. Rank product of preference orderings from A, B, and C above.

Because paragraphs thrown out in A, B and C are rank 0, at this step only paragraphs with feature matches and protein mentions remain. The resulting three rankings constitute the paragraph rankings used in the three runs submitted for the IPS subtask: 1, 2, and 3, respectively.

#### *Mapping of protein mentions to UniProt IDs*

To obtain the actual PPI pairs contained in the paragraphs of ranks 1, 2 and 3 described in the previous subsection, we had to convert the textual mentions obtained with ABNER to UniProt IDs. Protein and gene references identified using the ABNER system were mapped to UniProt IDs through exact matching with either a gene or a protein name occurring in SwissProt - considering both primary names and synonyms. UniProt version 8.2 was used for the mapping; this is not the most current version and could have resulted in missing relevant mappings. These mappings were then filtered using the reduced UniProt subset provided by BioCreative II. This process typically resulted in many UniProt IDs for the same ABNER protein mention, mostly because the same protein name maps to different UniProt IDs for different organisms. We therefore filtered the protein mention to include only UniProt ID mappings associated with organisms in the set of MeSH terms of a given article. Unfortunately, many of the articles listed several organisms in their MeSH terms, and so in that case our system returns several UniProt IDs for the same protein mention, resulting in the high recall observed (see Results [above]).

#### *Selection and ranking of protein-protein interaction pairs for IPS*

Finally, for the IPS task we returned all the combinations of protein pairs (UniProt accession numbers) occurring in the same sentence - for sentences included in the paragraphs of ranks 1, 2, and 3 above. For a given document (PMID), the rank of each PPI pair is the rank of the highest ranked paragraph in which the pair occurs in a sentence. We submitted three distinct rankings of PPI pairs according to the three ranks 1, 2, and 3 above. Because only paragraphs with feature matches and protein mentions remain after computing ranks 1, 2, and 3, we return a ranked list of all PPI pairs identified in every paragraph still in these three ranks.

#### *Protein mention feature expansion with proximity networks*

For the ISS subtask, we used only the PPI pairs that were obtained via the IPS rank 1 above. (We did not use IPS ranks 2 and 3, simply because we computed the sentence feature set too close to the challenge deadline to be able to use in ISS the PPI pairs identified via these two additional IPS ranks.) We used a version of a method we previously employed in the first BioCreative competition to obtain additional, contextualized features associated with protein mentions [9]. The assumption is that words that tend to co-occur in a given document, with the protein names identified in that document as PPI pairs, are useful as additional features. Furthermore, we assume that these expanded features are relevant in a specific document, but not necessarily in the whole corpus. To obtain these expanded features, we computed for each document a word proximity network based on a co-occurrence measure of proximity of stemmed words in paragraphs of that document, which we refer to as 'word paragraph proximity'. The weights of the edges of this network of words are given by the following equation:

$$wpp(w_i, w_j) = \frac{\sum_{k=1}^{m} (d_{k,i} \wedge d_{k,j})}{\sum_{k=1}^{m} (d_{k,i} \vee d_{k,j})} \tag{8}$$

Where $d_{i,j} \in \{0.1\}$ is an element of the document's paragraph × words relation $D$: $P \times W$; $P$ is the set of all $m$ paragraphs in a document, and $W$ is the set of all $n$ stemmed words also from that document. This yields a proximity network for each document, where the nodes are words $w_i$, and the edges are the $wpp(w_i, w_j)$ proximity weights.

Figures 10 and 11 depict an example word proximity network constructed from article 10464305 [24]. The title of this article is 'Stable association of 70-kDa heat shock protein induces latent multisite specificity of a unisite-specific endonuclease in yeast mitochondria'. It reports that the 50 kDa subunit of the endonuclease Endo.SceI of yeast mitochondria forms a stable dimer with the mitochondrial 70 kDa heat shock protein (mtHSP70), where the mtHSP70 is shown to act on the 50 kDa subunit with high specificity and activity. The subnetwork shown in Figure 11 shows a cluster of co-occurring words in this document, which captures much of this specific context. In this network, stemmed, overall corpus feature words such as 'protein', 'specif', 'dna', 'with', and 'reveal' are very near (high proximity to) more context-specific words

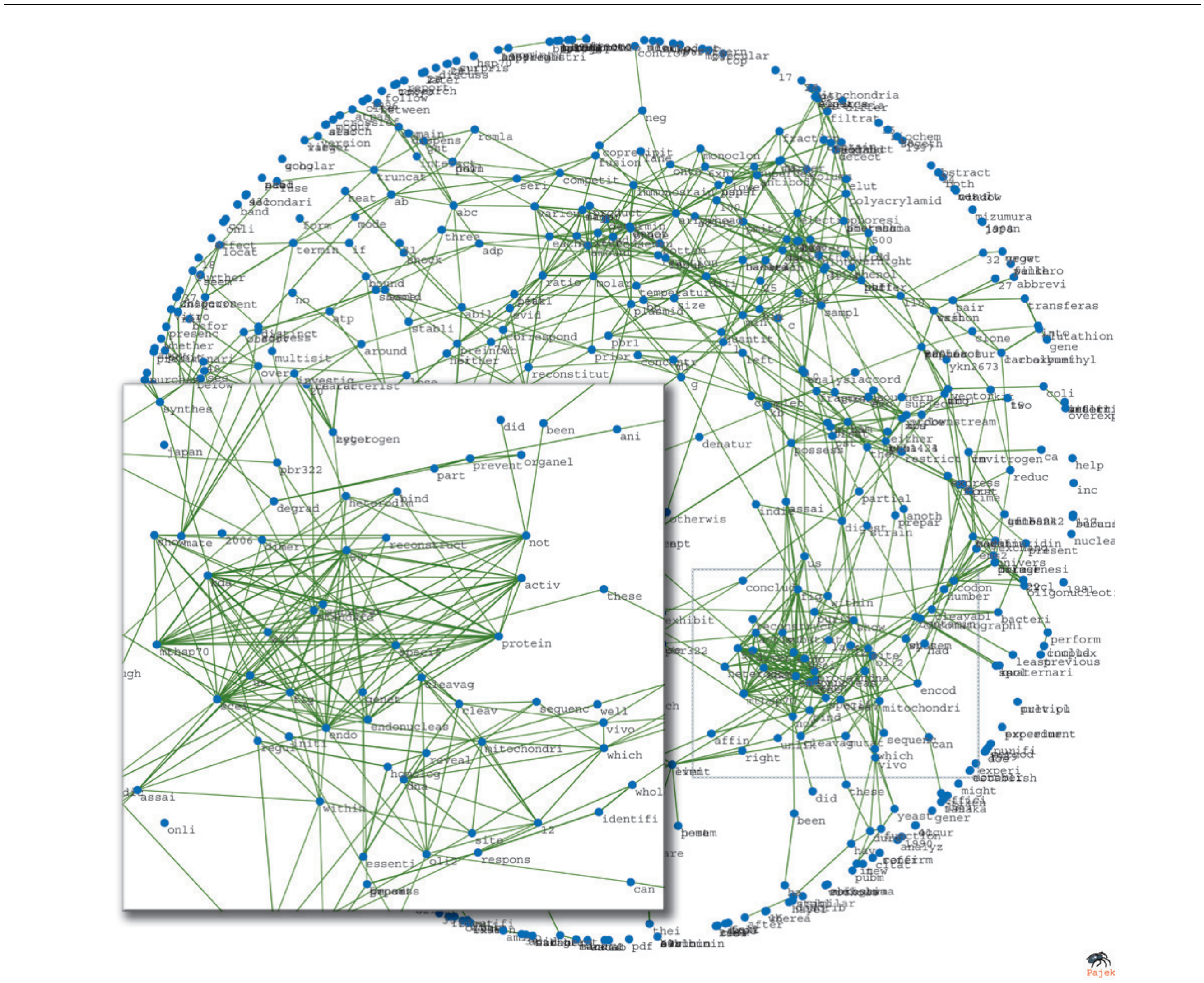

**Figure 10**

Word proximity network for document 10464305. Proximity network of 706 stemmed words produced from document 10464305 [24]. Showing only edges with proximity weights (formula 8) greater than 0.4. Inset detail showing cluster of highly associated words very related to the specific context of the article, whose title is 'Stable association of 70-kDa heat shock protein induces latent multisite specificity of a unisite-specific endonuclease in yeast mitochondria'. Plotted using Pajek such as 'mitochondri', 'mtHSP70', 'kda', 'endonuclease', and so on. This way, the more generic features extracted from the entire training data to detect protein interaction can be expanded with words that are specific to the context of the article, which can in principle improve the detection of the best sentences to describe protein interaction.

Next, for every PPI pair (obtained by IPS rank 1) occurring in a given document, we obtain the words closest to the protein labels in the document's proximity network. Notice that these protein labels are words identified by ABNER for the given PPI pair, and they should appear on the proximity network as regular nodes - unless stemming or other processing breaks them. For each protein pair we selected the five stemmed words (nodes) in the proximity network with largest minimum proximity to both protein names. These additional stemmed words were then added to the list of general features obtained from the training data, but only for the respective document. Therefore, each document contains general word features extracted from the entire corpus, plus five specific word features near to each PPI pair in the proximity network. The assumption is that these additional word features endow our method with additional context sensitivity.

### *Passage extraction and ISS submission*

From ranked paragraphs, we selected passages (sets of three sentences) containing a given PPI pair. Finally, we submitted three runs to the ISS subtask.

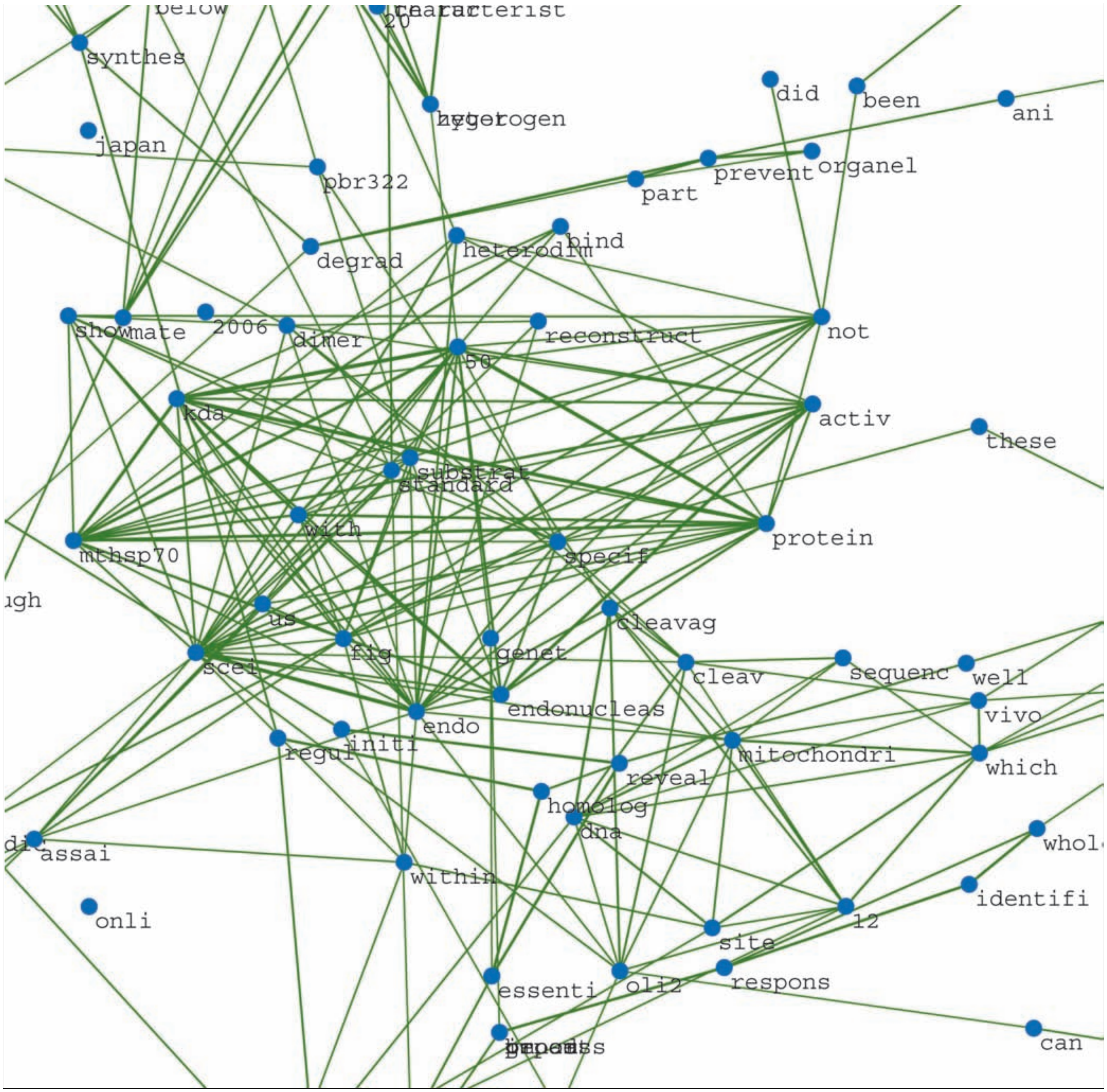

**Figure 11**
Detail of word proximity network for document 10464305. Proximity subnetwork of cluster of stemmed words produced from document 10464305 [24]. Showing only edges with proximity weights (Equation 8) greater than 0.4. This cluster shows highly associated words very related to the specific context of the article, whose title is 'Stable association of 70-kDa heat shock protein induces latent multisite specificity of a unisite-specific endonuclease in yeast mitochondria'. Plotted using Pajek.

1. Passages ranked by largest number of occurring word pair features.

2. Passages ranked by largest number of occurring word pair features, expanded with the extra, context-specific words extracted from the document's proximity network as described in the previous subsection.

3. Same as 2, with the addition of the following factor (100/[rank 1 from IPS submission]) to the number of features found in the passage.

## Abbreviations

ABNER, A Biomedical Named Entity Recognizer; AUC, area under the receiver operating characteristic curve; IAS, interaction article subtask; IDF, inverse document frequency; IPS, interaction pair subtask; ISS, interaction sentences subtask; MINT, Molecular Interactions Database; PIARE, Protein Interaction Abstract Relevance Evaluator; PPI, protein-protein interaction; SVD, singular value decomposition; SVD-UI, SVD with uncertainty integration; SVM, support vector machine; TN, true negative; TP, true positive; VTT, variable trigonometric threshold.

## Competing interests

The authors declare that they have no competing interests.

## Authors' contributions

AA-H produced all code necessary for preprocessing abstracts, computing training data partitions, and conducting the tests of the IAS task; he also participated in the development of the VTT algorithm and feature extraction in the IPS and ISS tasks. JK produced sentence features as well as general programming and analysis in the IPS and ISS tasks. AM programmed, tested, and participated in the development of the uncertainty integration (SVD-UI) method as well as ran ABNER for the IAS task; she also produced the proximity networks and programmed the feature expansion method used in the ISS task. PR conducted all portions of this work related to our SVM run as well as the SVM postchallenge analysis of data for the IAS task; he also extracted additional protein-relevant abstracts. AR computed and performed the SVD analysis in the IAS task. KV linked ABNER output to Uniprot data in the IPS task; she also participated in using ABNER in the IAS task. ZW participated in extracting features from full-text articles as well as general programming necessary for the IPS and ISS tasks. LMR was responsible for integrating the team and designing the experimental set up for all tasks; he also participated in the development of the VTT and SVD-UI methods in IAS, and developed the methods used in the IPS and ISS tasks, including the proximity network feature expansion.

## Acknowledgements

We would like to thank Santiago Schnell for graciously providing us with additional proteomics-related articles not containing PPI information. We would also like to thank the FLAD Computational Biology Collaboratorium at the Gulbenkian Institute in Oeiras, Portugal, for hosting and providing facilities used to conduct part of this research. It was at the collaboratorium that we interacted with Florentino Riverola, whose SpamHunting system inspired our approach to the IAS task, and who was most helpful in discussing his system with us. We are also grateful to Indiana University's Research and Technical Services for technical support. The AVIDD Linux Clusters used in our analysis are funded in part by NSF Grant CDA-9601632.

This article has been published as part of *Genome Biology* Volume 9 Supplement 2, 2008: The BioCreative II - Critical Assessment for Information Extraction in Biology Challenge. The full contents of the supplement are available online at http://genomebiology.com/supplements/9/S2.